\documentclass[fleqn,usenatbib]{mnras}

\usepackage{newtxtext,newtxmath}

\usepackage[T1]{fontenc}
\usepackage{soul}
\usepackage{graphicx}	
\usepackage{amsmath}	
\usepackage{booktabs}
\usepackage{longtable}
\usepackage{xcolor}
\usepackage{hyperref}

\newcommand{\hi}{\ion{H}{1}}
\newcommand{\cii}{\ion{C}{2}}
\newcommand{\ciii}{\ion{C}{3}}
\newcommand{\ciiin}{\ion{C}{3}}
\newcommand{\civ}{\ion{C}{4}}
\newcommand{\civn}{\ion{C}{4}}

\newcommand{\ovi}{\ion{O}{6}}
\newcommand{\ovii}{\ion{O}{7}}
\newcommand{\oviii}{\ion{O}{8}}
\newcommand{\siii}{\ion{Si}{2}}
\newcommand{\siiii}{\ion{Si}{3}}
\def\hi{{{\rm H}\,{\sc i}}}
\def\cii{{{\rm C}\,{\sc ii}~}}
\def\ciii{{{\rm C}\,{\sc iii}~}}
\def\ciiin{{{\rm C}\,{\sc iii}}}
\def\civ{{{\rm C}\,{\sc iv}~}}
\def\civn{{{\rm C}\,{\sc iv}}}

\def\ovi{{{\rm O}\,{\sc vi}~}}
\def\ovii{{{\rm O}\,{\sc vii}~}}
\def\oviii{{{\rm O}\,{\sc viii}~}}
\def\siii{{{\rm Si}\,{\sc ii}~}}
\def\siiii{{{\rm Si}\,{\sc iii}~}}
\newcommand{\ovin}{\ion{O}{6}}
\newcommand{\oviin}{\ion{O}{7}}
\newcommand{\oviiin}{\ion{O}{8}}
\def\ovin{{{\rm O}\,{\sc vi}}}
\def\oviin{{{\rm O}\,{\sc vii}}}
\def\oviiin{{{\rm O}\,{\sc viii}}}
\def\xmm{{\it XMM-Newton}}
\def\chandra{{\it Chandra}}

\title[All-sky z=0 dispersion measure]{\textcolor{black}{Empirical estimates of the} Galactic halo contribution to the dispersion measures of extragalactic fast radio bursts \textcolor{black}{using X-ray absorption}}

\author[Das et al.]{
Sanskriti Das,$^{1}$\thanks{E-mail: das.244@buckeyemail.osu.edu}
Smita Mathur,$^{1,2}$
Anjali Gupta,$^{1,3}$
Fabrizio Nicastro$^{4,5}$
and Yair Krongold$^{6}$
\\
$^{1}$Department of Astronomy, The Ohio State University, 140 West 18th Avenue, Columbus, OH 43210, USA\\
$^{2}$Center for Cosmology and Astroparticle Physics, 191 West Woodruff Avenue, Columbus, OH 43210, USA\\
$^{3}$Columbus State Community College, 550 E Spring St., Columbus, OH 43210, USA\\
$^{4}$Observatorio Astronomico di Roma - INAF, Via di Frascati 33, 1-00040 Monte Porzio Catone, RM, Italy\\
$^{5}$Harvard-Smithsonian Center for Astrophysics, 60 Garden St., MS-04, Cambridge, MA 02138, USA\\
$^{6}$Instituto de Astronomia, Universidad Nacional Autonoma de Mexico, 04510 Mexico City, Mexico
}

\begin{document}
\label{firstpage}
\pagerange{\pageref{firstpage}--\pageref{lastpage}}
\maketitle

\begin{abstract}
\noindent We provide an empirical list of the Galactic dispersion measure ($DM_{Gal}$) contribution to the extragalactic fast radio bursts along 72 sightlines. It is independent of any model of the Galaxy, i.e., we do not assume the density of the disk or the halo, spatial extent of the halo, baryonic mass content, or any such external constraints to measure $DM_{Gal}$. We use 21-cm, UV, EUV and X-ray data to account for different phases, and find that $DM_{Gal}$ is dominated by the hot phase probed by X-ray absorption. We improve upon the measurements of N(\oviin) and f$_{OVII}$ compared to previous studies, thus providing a better estimate of the hot phase contribution. The median $DM_{Gal}$=64$^{+20}_{-23}$ cm$^{-3}$ pc, with a 68\% (90\%) confidence interval of 33--172 (23--660) cm$^{-3}$ pc. The $DM_{Gal}$ does not appear to follow any trend with the galactic longitude or latitude, and there is a large scatter around the values predicted by simple disk$+$\textcolor{black}{spherical} halo models. Our measurements provide more complete and accurate estimates of $DM_{Gal}$ \textcolor{black}{independent} from the previous studies. We  provide  a  table  and  a  code  to  retrieve $DM_{Gal}$ for any  FRB  localized  in  the  sky.
\end{abstract}




\section{Introduction}
Fast radio bursts (FRBs) are bright (50 mJy--100 Jy) coherent pulses of emission at radio frequencies, with duration of order milliseconds or less \citep{Lorimer2007,Petroff2019a}. The intervening plasma through which the pulses travel imposes a refractive index that retards the group velocity as a function of frequency. This leads to a time delay 
($\Delta t$) between the highest ($\nu_h$) and lowest ($\nu_l$) radio frequencies of the pulse, quantified by the dispersion measure (DM):  $DM \propto \Delta t/(\nu_l^{-2} - \nu_h^{-2})$. The DM of an FRB at redshift $z$ is defined as $DM = \int \frac{n_e}{1+z} dl $, a line-of-sight integration of the free-electron number density of the intervening medium. Typically, the DM of FRBs are hundreds (and sometimes a few thousands) of cm$^{-3}$ pc \citep{Petroff2016}, which is too large to be explained by the free electrons in the interstellar medium (ISM) of the Milky Way \citep{Cordes2002,Dolag2015,Yao2017}. This indicates that FRBs are extragalactic. 

The extragalactic origin of FRBs makes it a promising tool to probe the otherwise invisible ionized intervening medium. Over the past decade, many uses of the DM of FRBs have been proposed, such as to study the cosmic reionization history, large-scale structure of the universe, cosmic proper distance measurements, baryon fraction of the intergalactic medium (IGM), and precision cosmology \citep[and references therein]{Zheng2014,Masui2015b,Yu2017,Li2019,Macquart2020}. 

The observed DM toward FRBs includes the DM of the host galaxy ($DM_{host}$), the intergalactic medium (IGM; $DM_{IGM}$), the Local Group, and the Milky Way. For any cosmological calculation using the DM of FRBs, it is necessary to know and remove the Galactic contribution, $DM_{Gal}$ from the total observed DM. By Galactic, we mean primarily the ISM in the disk and the circumgalactic medium (CGM) in the halo of the Milky Way. Because of the unknown spatial extent of the Galactic halo, the DM signatures of the Local Group and the Galaxy halo become observationally indistinguishable, broadly providing the $z\approx0$ value. 

The surveys searching for FRBs usually set a cutoff on the DM such that the DM of a detected FRB is larger than $DM_{Gal}$ \citep{Petroff2019a}. The DM$_{IGM}-z$ relation can be used to roughly estimate the redshift of an FRB \citep{Zheng2014,Li2019}, but to estimate $DM_{IGM}$ the knowledge of $DM_{Gal}$ is essential. Therefore, it is important to have a detailed understanding of the sky distribution of the Galactic DM to efficiently detect FRBs and to measure their distances, which again is instrumental for cosmological studies. 

\textcolor{black}{Because of the larger sky coverage than the Galactic sightlines, extragalactic sightlines are crucial for studying the IGM. Therefore, it is necessary to accurately estimate the contribution of Galactic halo along the extragalactic sightlines.}
\cite{Yamasaki2020} prescribed a disk-like plus a spherical halo density model from the X-ray emission measure of the Galactic halo along $>100$ sightlines \textcolor{black}{\citep[see also][]{Keating2020}} and predicted the Galactic DM contribution based on that model. \textcolor{black}{This model is complementary to the previous models \citep[e.g.,][]{Cordes2002,Yao2017} around the disk where the hot gas and other cooler phases have comparable contribution,} but at high galactic latitude ($|b| \geqslant 20^\circ$) this model is a significant improvement over previous models which ignored the halo component for simplicity.  

As emission measure (EM) is proportional to the density squared (EM=$\int
n_en_p dl$), it is not possible to retrieve the dispersion measure from the EM without constructing a density model. Often, such models depend on many parameters including the spatial extent of the Galactic halo, the baryon fraction in the halo and the virial mass of Milky Way. However, none of these quantities are well-constrained and the spatial extent varies wildly all over the sky \citep[see][for details]{Boylan2013,Gupta2012,Gupta2017}. This leads to a huge systematic uncertainty which usually surpasses the statistical uncertainty of the emission measurements from which the density model is constructed. 

On the other hand, the column density ($N_X = \int n_X dl$) from absorption analyses can be directly converted to the dispersion measure assuming some ionization condition. \cite{Prochaska2019} used the column densities of \ovii K-$\alpha$ lines from \cite{Fang2015} to calculate the DM contribution of the hot Galactic halo and constrained their proposed halo model accordingly. The equivalent widths of the \ovii K-$\alpha$ lines indicate that many lines are saturated but not damped. The spectral resolution of the Reflection Grating Spectrometer (RGS) of \xmm~is not good enough to resolve the \ovii line and obtain the velocity width. Therefore, the Voigt profile fitting, as has been done in \cite{Fang2015}, might not be an accurate way to obtain the column density of these lines.  Instead, the equivalent widths of the \ovii K-$\alpha$ and \ovii K-$\beta$ lines can be combined to constrain the column density and the velocity width \citep[e.g.,][]{Nicastro2002,Williams2005, Gupta2012,Nicastro2016a,Gupta2017}; this is our approach for calculating the \ovii column densities in this paper. 

We provide an empirical estimation of the DM contribution of the Galactic disk and halo from the X-ray absorption analyses. For completeness, we have considered other phases, although those are not the primary contributors. Instead of constructing a density model, we provide the DM along the observed sightlines. It is a more appropriate representation of the Galactic DM than the previous estimates.  

This paper is organized as follows. In section \ref{sec:analys} we discuss the steps to calculate the Galactic dispersion measure. In section \ref{sec:result} we show how the Galactic dispersion measure contribution is distributed over the sky  and compare it with previous models. Finally in section \ref{sec:summary} we summarize the result. 

\section{Analysis}\label{sec:analys}
\noindent The ISM and CGM are multiphase, so we accumulate the data from the literature in different wavelengths probing different phases. We do not assume any density model; we convert the observed quantities to the Galactic dispersion measure with the  assumptions standard in the respective fields. In fact, by using more observables, we make fewer assumptions (\S2.2).  We further note that the DM models based on the emission of the hot phase ignore all other phases, and the previous absorption based study includes some (but not all) of the cooler phases. As such, our study is more comprehensive and more accurate compared to previous studies.

The total Galactic dispersion measure (DM) contribution is be a combination of the disk and the halo in four different phases: 
\begin{equation}
    DM_{Gal} = DM_{cold} + DM_{cool} + DM_{warm} + DM_{hot} 
\end{equation}
Here, ``cold'' refers to  $\approx 10^4$K gas which is predominantly neutral, ``cool'' is $10^{4-5}$K mildly ionized gas, ``warm'' is for $\approx 10^{5-5.5}$ K gas probed by Li-like ions (primarily \ovin), and hot refers to  $\geqslant 10^6$K gas probed by H- and He-like ions, e.g., \ovii and \oviii \citep{Tumlinson2017}.    

We obtain DM$_{cold}$ from the 21-cm \hi~emission measurement at $z=0$ \citep{Bekhti2016}
using the following equation: 
\begin{equation}
    DM_{cold} = 6.5 cm^{-3} pc \Big(\frac{N_{HI}}{10^{21}cm^{-2}}\Big)\Big(\frac{x_e}{0.02}\Big)
\end{equation}
Here, $x_e = \frac{n_e}{n_H}$ is the electron fraction, which is typically 0.02 for a $\approx 10^4$K gas \citep{Draine2011}. 

The cool gas is probed by singly/doubly ionized gas (e.g., \siii and \siiii ions). By assuming that the element is in only two ionization states, we obtain DM$_{cool}$ from the column densities of \siii and \siiii ions (also \cii and \civ ions) using the following equation:
\begin{equation}
\begin{split}
    DM_{cool} = 5.3 cm^{-3} pc  \Big(\frac{N_{SiII}+N_{SiIII}}{2\times 10^{13.4}cm^{-2}}\Big) \\
    \Big(\frac{A_{Si,\odot}}{3.2 \times10^{-5}}\Big)^{-1} \Big(\frac{Z}{0.1Z_\odot}\Big)^{-1}
\end{split}    
\end{equation}
Here, we scale with respect to the median column density of \siii and \siiii in the intermediate and high velocity absorbers in the Galactic halo \citep{Richter2017}, corrected by a factor of 2 to account for the low velocity absorbers \citep{Zheng2015}. $A_{Si,\odot}$ is the solar abundance of silicon \citep{Asplund2009}. The typical metallicity is taken to be 0.1 Z$_\odot$ \citep{Wakker2001}. A similar calculation with \cii and \civ from \cite{Richter2017} yields a DM value of 3.2 cm$^{-3}$ pc.

 The assumption of all \siii and \siiii coming from the same medium might not generally be true. The cool phase is photo-ionized, and the uncertainties related to photo-ionization are large. The column densities of \siii and \siiii and their ratios  span over an order of magnitude over the whole sky \citep{Richter2017}, indicating the complex thermal and ionization structure. For Carbon in the cool phase, observed absorption lines are from \cii and \civn, but not \ciiin, but in the photoionized gas \ciii must exist together with \cii and \civn. This will make DM$_{cool}$ based on Carbon, similar than from Si. 

{It should be noted that all of the \hi~measured in 21-cm might not come from a predominantly neutral medium. If \hi~comes from an ionized medium, DM$_{cold}$, the DM in the cold phase, would be lower. 
One would generally expect denser mediums (i.e., with (N(\hi) $\geqslant 10^{20}$ cm$^{-2}$) to be more shielded and hence predominantly neutral, while smaller N(\hi) values can come from a partially ionized medium. Based on the typical N(\hi) values in the cool phase, we estimate an approximate DM. The average ionization fraction of \hi, $f_{HI} = 0.3$ in the cool phase}  \citep{Lehner2011,Putman2012,Richter2017}. The median N(\hi) in \cite{Bekhti2016} is N(\hi) = $4.3\times 10^{18}$ cm$^{-2}$. {The DM for this median N(\hi) and average $f_{HI}$ would be 6.5 cm$^{-3}$ pc}, including the correction factor of 2 to account for the low velocity gas \citep{Zheng2015}. {This is comparable with DM$_{cool}$ obtained from silicon and carbon lines using equation 3}, as was also found by \cite{Prochaska2019}. Therefore, between the cold and cool phases, the sightlines with high N(\hi) are likely to have a higher DM$_{cold}$ and the smaller N(\hi) would contribute higher DM$_{cool}$. As the DM calculated from metal lines and \hi~are comparable, we consider the DM calculated from metal lines as the bulk estimate of DM$_{cool}$. To avoid double counting, we do not include the cool phase in the final calculation. If along a given sightline DM$_{cold} \ll$ DM$_{cool}$ indicating that DM is dominated by the cool phase rather than the cold phase, DM$_{cool}$ might be added to the final calculation of DM$_{Gal}$. 

We estimate the DM contribution of the warm phase using the following equation:
\begin{equation}
\begin{split}
    DM_{warm} = 4.4 cm^{-3} pc \Big(\frac{N_{OVI}}{2\times10^{14.3}cm^{-2}}\Big) \Big(\frac{f_{OVI}}{0.2}\Big)^{-1}\\\Big(\frac{A_{O,\odot}}{4.9\times10^{-4}}\Big)^{-1} \Big(\frac{Z}{0.3Z_\odot}\Big)^{-1}
\end{split}    
\end{equation}
Here, we scale with respect to the median column density of \ovi in the intermediate and high velocity absorbers in the Galactic halo \citep{Sembach2003}, corrected by a factor of 2 to account for the low velocity absorbers \citep{Zheng2015}. $A_{O,\odot}$ is the solar abundance of oxygen \citep{Asplund2009}. The typical metallicity is taken to be 0.3 Z$_\odot$, the median metallicity of the warm CGM of L* galaxies in the COS-Halos sample \citep{Prochaska2017}. We adopt the ionization fraction of $f_{\small{\hbox{\ovi}}}=0.2$, maximum for the collisional ionization equilibrium. 

\begin{figure*}
    \centering
    \includegraphics[trim=15 3 50 30,clip,  scale=0.475]{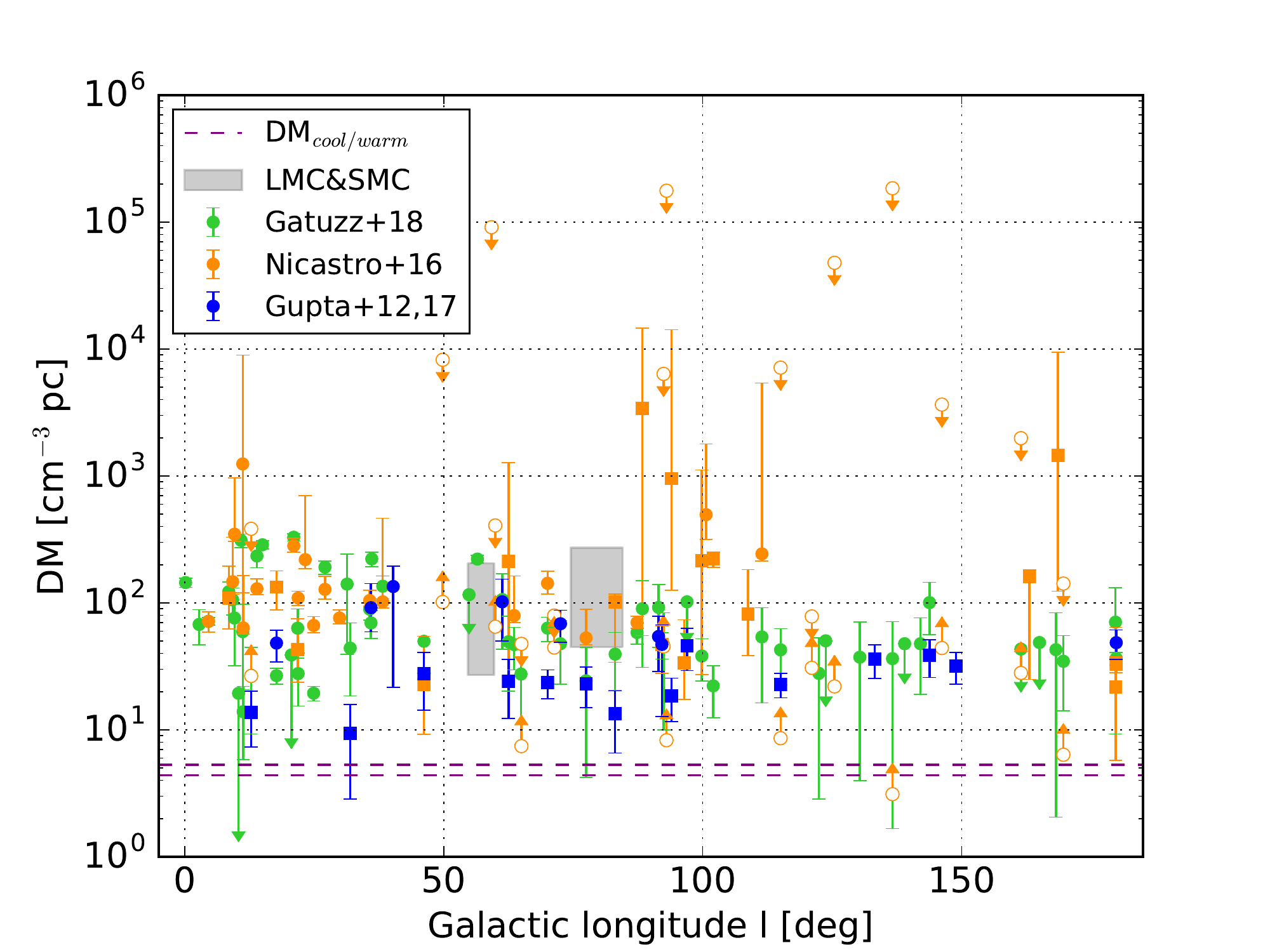}
    \includegraphics[trim=15 3 45 30,clip,scale=0.475]{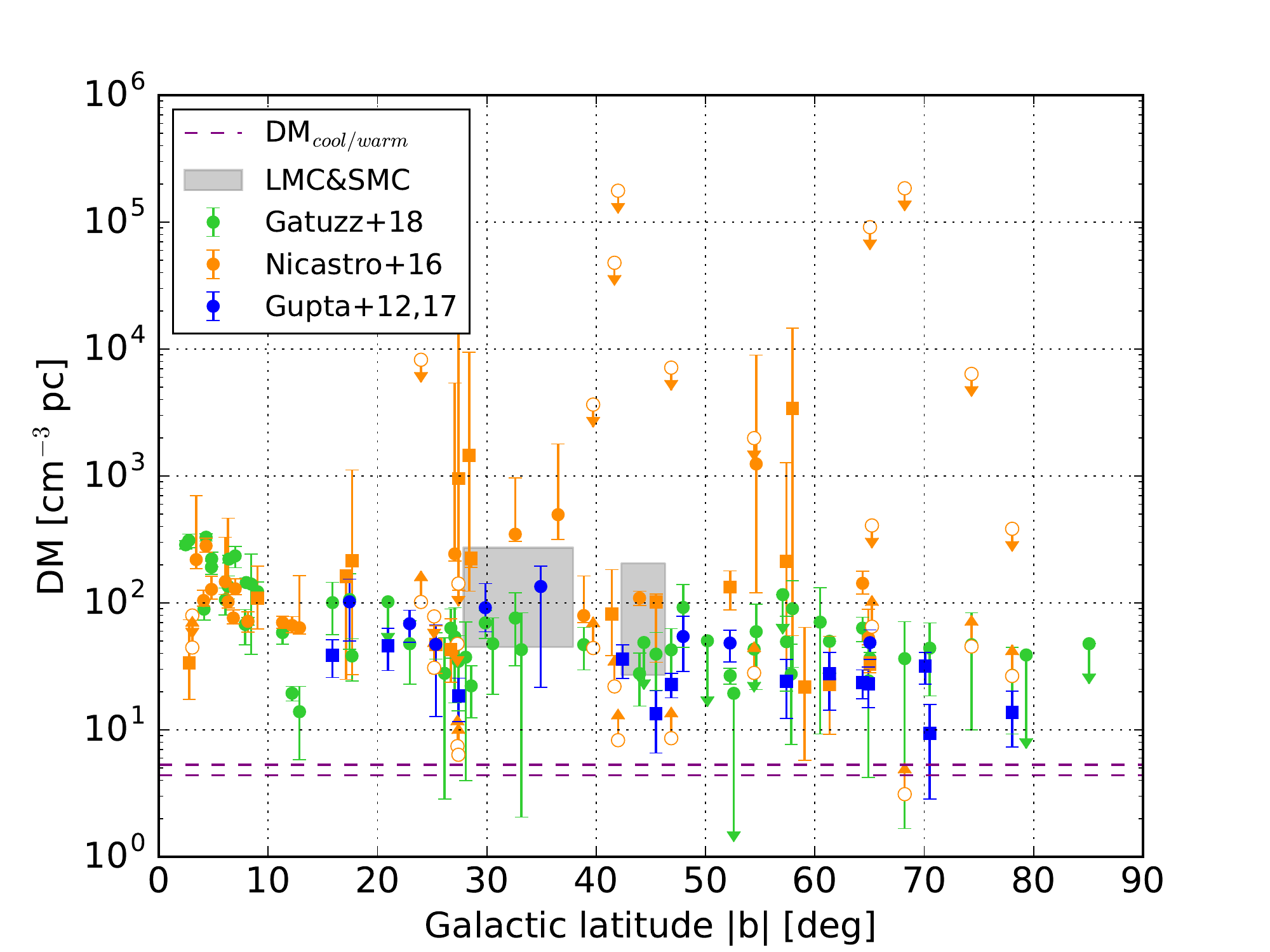}
    \includegraphics[trim=15 3 50 40,clip,  scale=0.475]{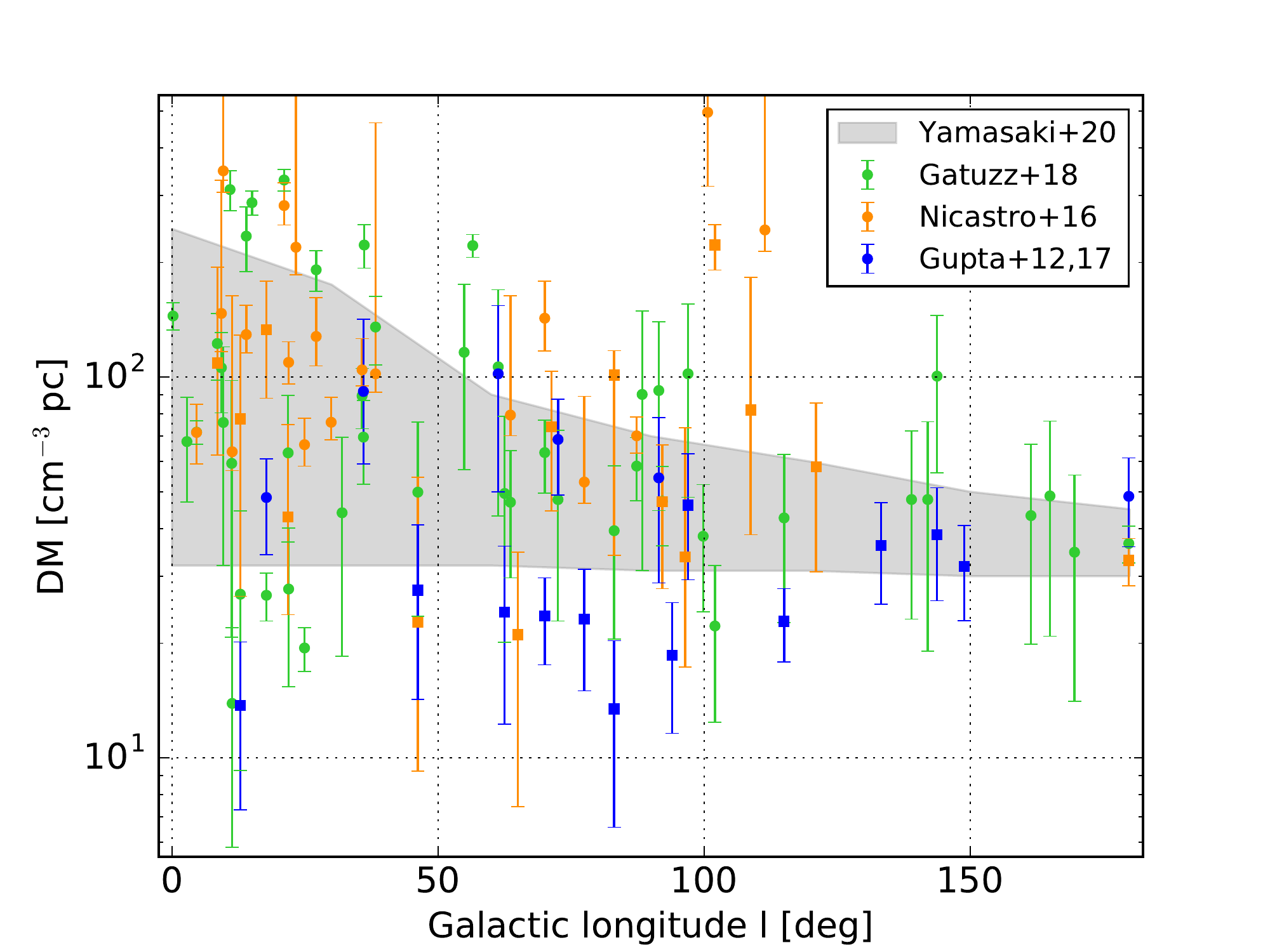}
    \includegraphics[trim=15 3 45 40,clip,  scale=0.475]{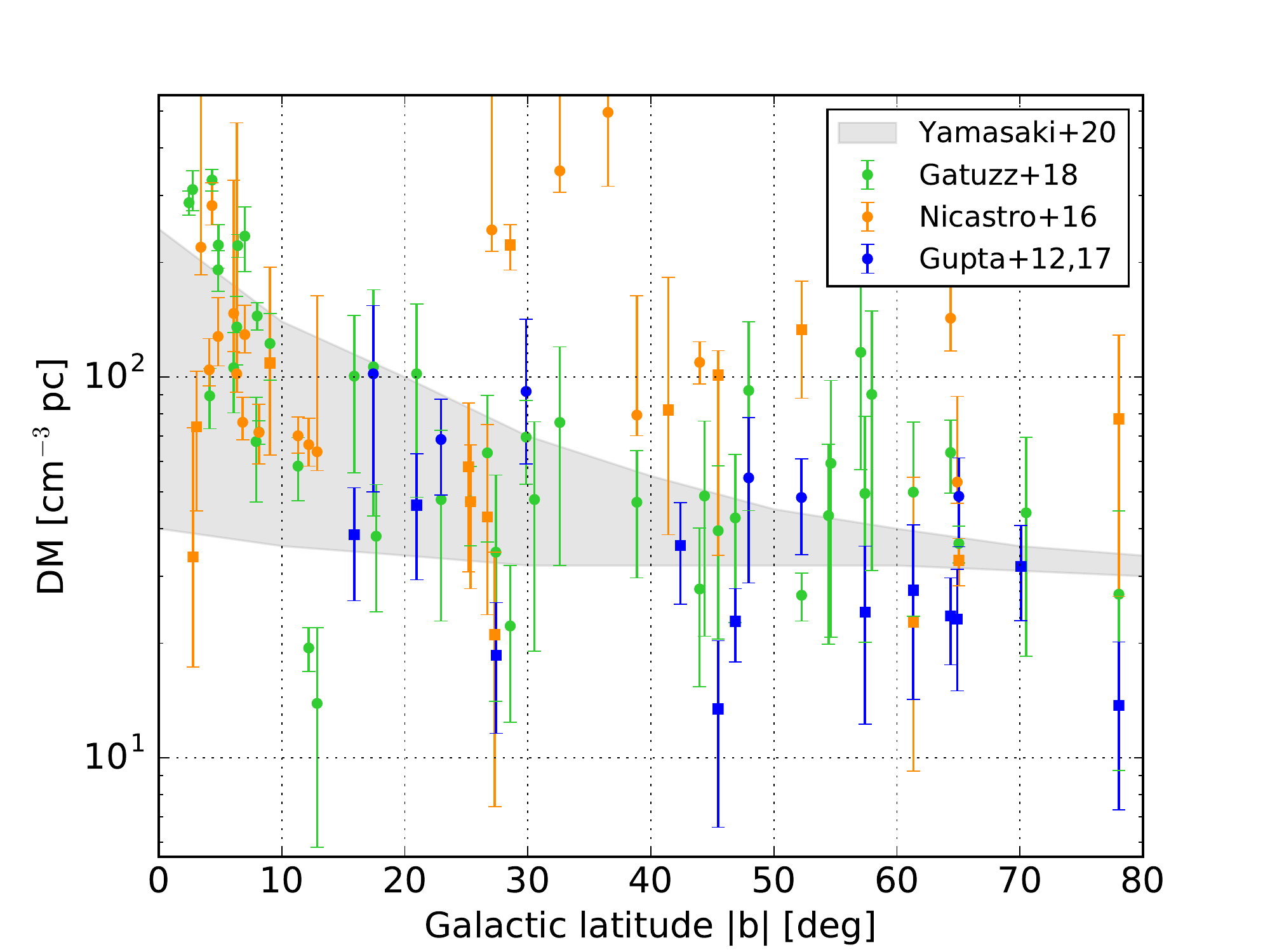}
    \caption{Dispersion measure (DM) contribution of the Galactic disk and halo as a function of galactic longitude (left) and latitude (right). The data points are derived from the 21-cm and X-ray absorption measurement of the cold \citep{Bekhti2016} and hot  \citep{Gupta2012,Nicastro2016a,Gupta2017,Gatuzz2018} Galactic halo. The filled circles denote the sightlines where both \ovii and \oviii lines were detected, the unfilled circles are where upper limits of \oviii were reported, squares denote the sightlines where the value of \oviii line was not reported and $f_{\hbox{\oviin}}$ is assumed to be 1 to calculate DM. Top: The gray regions denote the range of the pulsar dispersion measure towards LMC and SMC \citep[and references therein]{Ridley2013}. The horizontal dashed line is the median of DM contribution of the cool and warm phases \citep{Richter2017,Sembach2003}. Bottom: The gray region is the predicted Galactic DM contribution modeled from the emission measure of hot Galactic halo \citep{Yamasaki2020}.}
    \label{fig:1d}
\end{figure*}

To calculate the DM contribution of the hot phase, we use \ovii K-$\alpha$, \ovii K-$\beta$ and \oviii Ly-$\alpha$ lines. Previous studies of DM assumed the ionization fraction of \oviin, $f_{OVII}$=1, and did not use the \ovii K-$\beta$ line to estimate N(\oviin). Unlike them, we use both \ovii K-$\alpha$ and K-$\beta$ lines to calculate N(\oviin), and the ratio of N(\oviiin) and N(\oviin) to calculate $f_{OVII}$ \citep[see also][]{Nicastro2002,Williams2005,Gupta2012}.  This is the main advancement in this paper compared to earlier DM estimates. 

DM$_{hot}$ is calculated by two methods. First, we calculate the DM from direct \ovii (and \oviiin, if available) line measurement by \cite{Gupta2012,Nicastro2016a,Gupta2017} using the following equation:
\begin{equation}
\begin{split}
    DM_{hot} = 83.7 cm^{-3} pc \Big(\frac{N_{OVII}}{10^{16.5} cm^{-2}}\Big)\Big(\frac{f_{OVII}}{1}\Big)^{-1}\\\Big(\frac{A_{O,\odot}}{4.9\times10^{-4}}\Big)^{-1} \Big(\frac{Z}{0.3Z_\odot}\Big)^{-1}
\end{split}
\end{equation}
Because \ovii K-$\alpha$ line in the CGM is often saturated but not damped, the column density cannot be directly estimated from the equivalent width (EW) of \ovii K-$\alpha$ by fitting it with a Gaussian/Voigt profile. For a given EW, the column density N in the flat portion of the curve-of-growth is a function of the velocity width $b$ \citep[N$ \propto b\times e^{b^{-2}}$,][]{Draine2011}. The intersection of the N vs. $b$ contours for the EWs of \ovii K-$\alpha$ and K-$\beta$ lines corresponds to the best-fitted column density and the velocity width of \oviin. The uncertainty is obtained from the range allowed by the confidence intervals of both contours at the best-fitted value. \oviii is optically thin, so N(\oviiin) is linearly proportional to EW(\oviiin).

If both \ovii and \oviii are detected or an upper limit exists along a sightline, the temperature (or its upper limit) of the gas is calculated from their column density ratio, $\frac{N_{OVIII}}{N_{OVII}}$, assuming that the gas is in collisional ionization equilibrium (CIE). The ionization fraction of \oviin, $f_{OVII}$ at that temperature is used to calculate DM$_{hot}$. Along some sightlines, the measurement of the \oviii line is not reported. This is either because \oviii line was too weak to obtain a measurement of the column, or \oviii could not be studied due to instrumental features at that wavelength. In these cases, we assume $f_{OVII}=1$\footnote{The maximum ionization fraction of \ovii in CIE is $\approx0.9$. But the measurement uncertainty in the column density of \ovii is much larger than this uncertainty in the ionization fraction. This validates the assumption of $f_{OVII}=1$.}. As metallicity cannot be measured in X-ray absorption due to the lack of a hydrogen line, we adopt the same metallicity in the hot and warm phases, assuming that the warm phase forms by cooling from the hot phase. 

The second method is based on the N(H) values of the hot phase estimated by \cite{Gatuzz2018}. There, the oxygen lines have been fitted using hybrid ionization modeling\footnote{The photo-ionization parameter is negligibly small in their model. Therefore, effectively, the model is collisional ionization.}, for a constant temperature of 10$^{6.3}$K. We take N(H) in the hot phase for solar metallicity from \cite{Gatuzz2018}\footnote{we discard the sightlines with an upper limit of N(H)}, and convert it to DM$_{hot}$ using
\begin{equation}
    DM_{hot} = 85.6 cm^{-3} pc \Big(\frac{N_{H}}{10^{20.3}cm^{-2}}\Big)\Big(\frac{Z}{0.3Z_\odot}\Big)^{-1}
\end{equation}
The N(H) of the cold phase was part of their model, whose value is not necessarily the same as the 21-cm measurement by \cite{Bekhti2016}. Therefore, we calculate DM$_{cold}$ of the \cite{Gatuzz2018} sightlines from their estimated N(H) of the cold phase. Thus the total DM$_{Gal}$ from \cite{Gatuzz2018} is ionization model-based, while other estimations \citep{Gupta2012,Nicastro2016a} are empirical.  

As can be seen from equations 2-6, The typical DM contribution of the hot phase ($>80$) exceeds the typical DM contribution of any other phase ($<8$) by almost an order of magnitude. Therefore, inclusion of those phases does not significantly affect the total Galactic DM contribution. Nonetheless, we add the contribution of the cold phase, because:\\
1) unlike the cool and warm phases, the 21-cm data is available along all the sightlines where X-ray data are available,\\
2) the DM contribution from the cool and warm phases are complicated by the uncertainties related to photo-ionization. The calculation in the cold phase, however, is straightforward, and \\
3) the relative contribution of the cold and hot phase may vary wildly over the sky because of the known anisotropy of the hot phase in the Galactic halo \citep{Henley2010,Gupta2012,Gupta2017,Nakashima2018}. 

\section{Results and discussion}\label{sec:result}
\noindent We tabulate the Galactic DM from the hot and the cold phases in tables \ref{tab:gupta}, \ref{tab:nicastro} and \ref{tab:gatuzz}. At smaller galactic latitudes, the cold phase has a significant contribution from the disk, and the contribution from the hot and the cold phases are comparable. At higher galactic latitudes ($b>20^\circ$) the contribution from the cold phase is negligible compared to the hot phase. 

We show the Galactic dispersion measure as a function of the galactic coordinates in Figure \ref{fig:1d}. We also plot the median DM contribution of the cool and warm phases to show that their values are negligible compared to the DM from the hot (and cold) phases (Figure \ref{fig:1d}, top). The pulsar dispersion measure toward LMC and SMC \citep{Ridley2013} are comparable with the DM values we obtain, validating the assumption of the 0.3 Z$_\odot$ metallicity in the hot phase.

The DM profile from the density model of \cite{Yamasaki2020} is shown for comparison {(Figure} \ref{fig:1d}, {bottom)}. The range of DM values at a given galactic latitude (longitude) is similar  to the range of DM values over the entire range of galactic longitude (latitude). On average, the model does a pretty good job of predicting the Galactic DM. However, {the exact DM value along many sightlines deviate significantly from the predicted profile, both at small and large (l,b). This shows that not even all the disk dominated sightlines can be explained by the disk+spherical halo model.} As the measurement of FRB DM is sightline specific, the estimation of Galactic DM and the cosmological calculations following that can be drastically incorrect if we use an average value. Therefore, the empirical values we present here are more accurate than the previously modeled values. 

\begin{figure}
    \centering
    \includegraphics[trim=20 4 45 30,clip,scale=0.45]{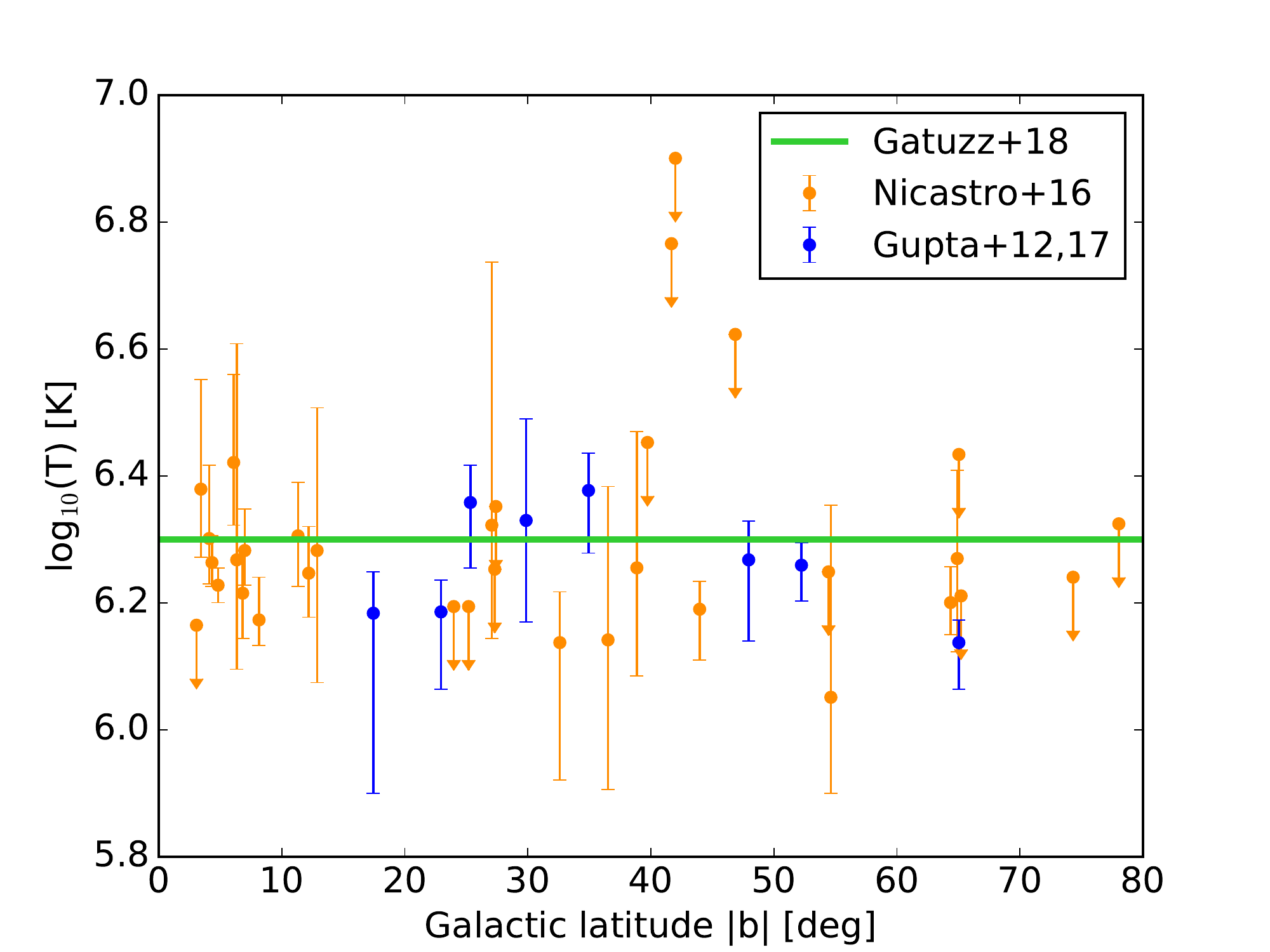}
    \caption{The temperature of the hot Galactic halo as a function of galactic latitude. The horizontal line corresponds to T $=10^{6.3}$ K which was assumed to be constant for all sightlines in $\hbox{\protect\cite{Gatuzz2018}}$. The data points are derived from the ratio or upper limit of the ratio of N(\oviiin) and N(\oviin) \citep{Gupta2012,Nicastro2016a} under the assumption of collisional ionization equilibrium.}
    \label{fig:T}
\end{figure}

Most of the sightlines have been observed with multiple instruments (\chandra~and \xmm) and/or multiple methods (direct measurement of absorption lines or ionization modeling). The DM estimates along some of the sightlines are not consistent with each other within error. This might partially be due to the assumption about the temperature (Figure \ref{fig:T}). For simplicity, \cite{Gatuzz2018} assumed a constant temperature of 10$^{6.3}$ K for all the sightlines. While this assumption is generally reasonable, the temperature obtained from the ratio of N(\oviiin) and N(\oviin) along every sightline is not necessarily the same. Because the ionization fraction of \ovii changes sharply around 10$^{6.3}$\,K \textcolor{black}{($f_{OVII}$ drops from 0.85 at 10$^{6.2}$\,K to 0.3 at 10$^{6.4}$\,K)}, the DM estimation is very sensitive to the temperature of the hot component. On the other hand, the ionization model might have a better continuum (and hence absorption line) estimation than the direct line measurements. The model simultaneously takes into account of multiple phases, including the absorption lines of multiple elements in addition to \ovii and \oviiin. Therefore, we do not have any particular reason to prefer one method over the other.  

\begin{figure}
    \centering
    \includegraphics[trim=35 30 55 45,clip,scale=0.5]{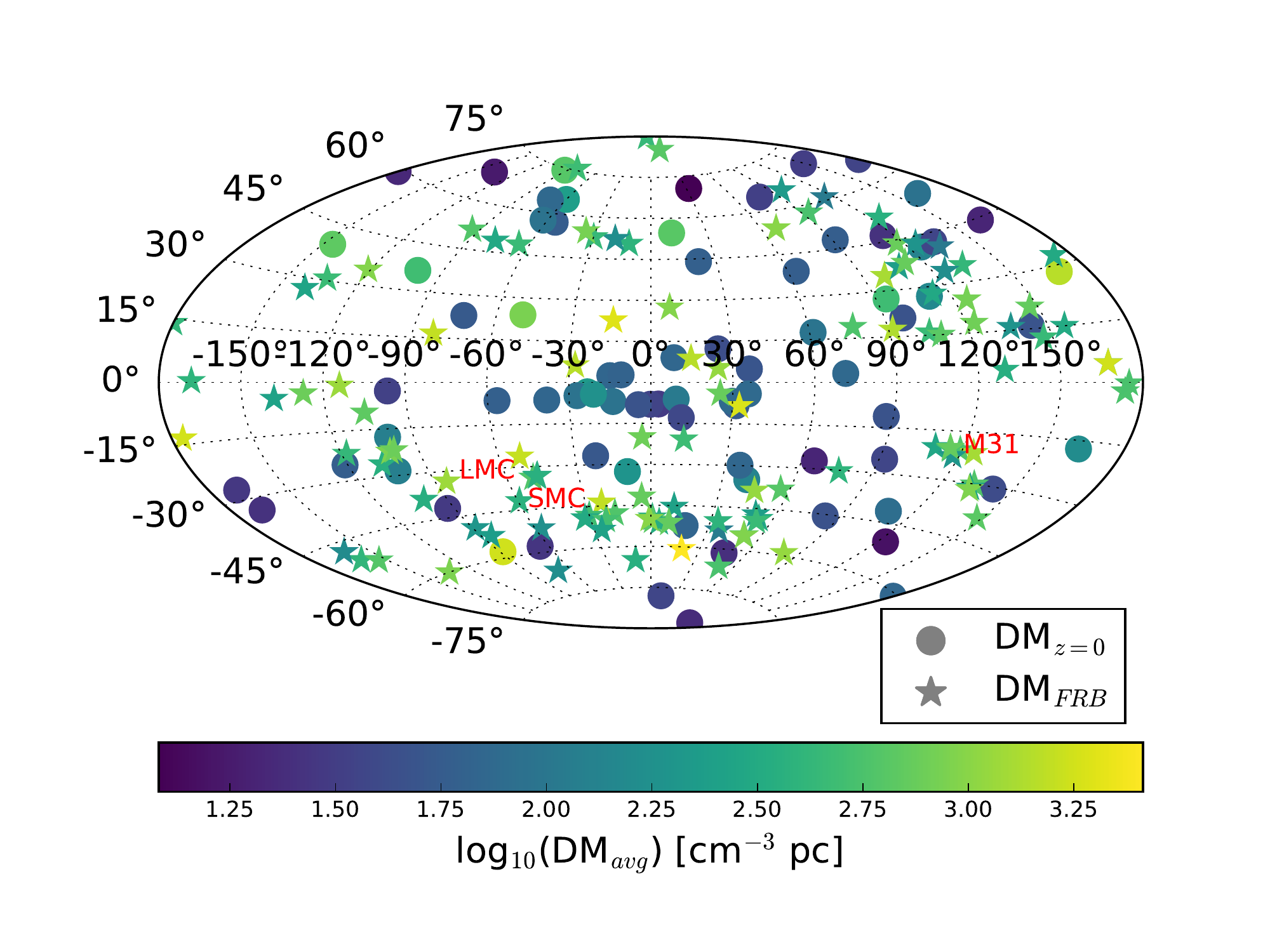}
    \caption{The average Galactic dispersion measure (DM) contribution obtained from multiple methods (direct measurement of \ovii and \oviii lines vs. hybrid ionization modeling) and/or instruments (i.e., \chandra~and \xmm) in the Aitoff projection of galactic coordinates. The DM of the  extragalactic FRBs observed in the past decade are plotted for comparison. The symbols are color-coded with the log$_{10}$ of DM; the filled circles are for the Galactic DM and the stars are for the FRB DM.}
    \label{fig:2d}
\end{figure}

The DM$_{Gal}$ does not have any trend with either of the galactic coordinates (Figure \ref{fig:1d}). There is more than two orders of magnitude scatter in the values of the DM$_{Gal}$. This shows that the geometrically ordered structures like the spherical halo and the disk might not explain the observation well. As DM$_{Gal}$ is dominated by DM$_{hot}$, this pattern of DM reflects the characteristics of the hot Galactic halo. This is consistent with the X-ray emission and absorption analyses which also report the hot Galactic halo to be inhomogeneous and anisotropic \citep{Gupta2012,Henley2013,Gupta2014,Gupta2017,Nakashima2018}.

\begin{table}
    \centering
    \caption{Correlation test. \textcolor{black}{The number of sightlines in each segment are mentioned in parentheses}}
    \begin{tabular}{ccc}
    \toprule
    Correlation with & $\tau$  & $p$ \\
    \midrule
    \multicolumn{3}{c}{All sightlines \textcolor{black}{(72)}} \\ \midrule
    $b$  & 0.059  & 0.460 \\
    $l$  & -0.044 & 0.586 \\
    \midrule
    \multicolumn{3}{c}{Extra-galactic: $|b|> 20^\circ$ \textcolor{black}{(46)}} \\ \midrule
    $b$ &  -0.005  & 0.962 \\
    $l$ &  -0.080   & 0.432 \\ 
    \midrule
    \multicolumn{3}{c}{Off-center: $ 20^\circ < l < 340^\circ$ \textcolor{black}{(59)}} \\ \midrule
    $b$ &  0.026  & 0.768 \\
    $l$ &  -0.070   & 0.436 \\ 
    \midrule
    \multicolumn{3}{c}{Northern hemisphere: $b> 0^\circ$ \textcolor{black}{(35)}} \\ \midrule
    $b$ &  -0.150  & 0.206 \\
    $l$ &  -0.106   & 0.371 \\ \midrule
    \multicolumn{3}{c}{Southern hemisphere: $b < 0^\circ$ \textcolor{black}{(37)}}\\\midrule
    $b$ &  0.162  & 0.158 \\
     $l$ &  -0.042   & 0.714 \\ \bottomrule
    \end{tabular}
    \label{tab:stat}
\end{table}

In Figure \ref{fig:2d} we show the sky distribution of Galactic DM contribution along with the DM of all FRBs discovered over past decade\footnote{The details of these FRBs are avaiable at \url{http://www.frbcat.org/}}.
For the sightlines observed multiple times using different instruments and/or analyzed in different methods, we plot the average DM. We also calculate the maximum DM along each sightline (Table \ref{tab:all}). The average DM should be the best estimate, while the maximum Galactic DM would provide a lower limit on the DM of the IGM.

As can be seen in Figure \ref{fig:1d}, the distribution of DM$_{Gal}$ in the sky does not have any orderly pattern (Figure \ref{fig:2d}). Two sightlines far apart can have similar DM$_{Gal}$, while close-by sightlines show a scatter in DM$_{Gal}$. We do not find any systematic increase in DM$_{Gal}$ toward the direction of M\,31 ($l=121.17^\circ, b=-21.57^\circ$), indicating that the halo of M\,31 might not have a significant contribution to the measured DM$_{Gal}$. Thus, interpolation of DM$_{Gal}$ values might not be the correct approach due to its non-monotonic behavior as a function of the galactic coordinates. This shows how complex the distribution of density, spatial extent, temperature and ionization state of the Galactic disk and halo are, and how challenging the modeling is to explain the details of the multi-wavelength observations.

\subsection{Statistical considerations}
We perform a Kendall's $\tau$ test to verify any correlation between DM$_{Gal}$ and the galactic coordinates (Table \ref{tab:stat}). We do not include the lower limits and upper limits in this test. The value of $\tau$ = +1/-1/0 implies a perfect positive/negative/null correlation, and the p-value is the probability of a null correlation. The all-sky distribution of DM$_{Gal}$ has a $|\tau| <0.1$ and a $\approx50\%$ probability of any correlation with $l$ or $b$. The lack of correlation between DM$_{Gal}$ and $b$ becomes more prominent when the extragalactic sightlines ($|b|>20^\circ$) or off-center sightlines ($20^\circ < l< 340^\circ$) are considered; the probability of a null correlation enhances to 96\% and 77\%, respectively. This shows why the disk model is not a good representation of the DM$_{Gal}$ distribution. The lack of strong anti-correlation between DM$_{Gal}$ and $l$ indicates that the halo is not isotropic, and hence, a spherical model might not be appropriate either. The correlations (or the lack there of) are not exactly similar in the two hemispheres. DM$_{Gal}$ in the northern hemisphere shows a weak anti-correlation with $b$, while the southern hemisphere shows a weak positive correlation. The probability of a null correlation with $l$ is higher (71\%) in the southern hemisphere than the northern hemisphere (37\%). These asymmetries are difficult to account for in the geometric density models. Once again, the empirical estimates are better. 

Our correlation coefficients (Table \ref{tab:stat}) are based on X-ray absorption studies along 72 sightlines. We compare these with the coefficients from \cite{Henley2013} based on X-ray emission measures (EM) along 110 sightlines. The EM distribution did not show any dependence on $|b|$ in either hemispheres, but there was significant ($p\ll 1$) anti-correlation with $l$ in the southern hemisphere. This is different from our $DM_{Gal}$ distribution. As the emission is dominated by denser regions, the disparity of correlations between EM and DM$_{Gal}$ distribution indicates that the hot gas probed in emission and absorption might not be the same. This also adds to the reasons for using absorption analyses for DM measurements.

\begin{figure}
    \centering
    \includegraphics[trim=40 2 45 30,clip,scale=0.5]{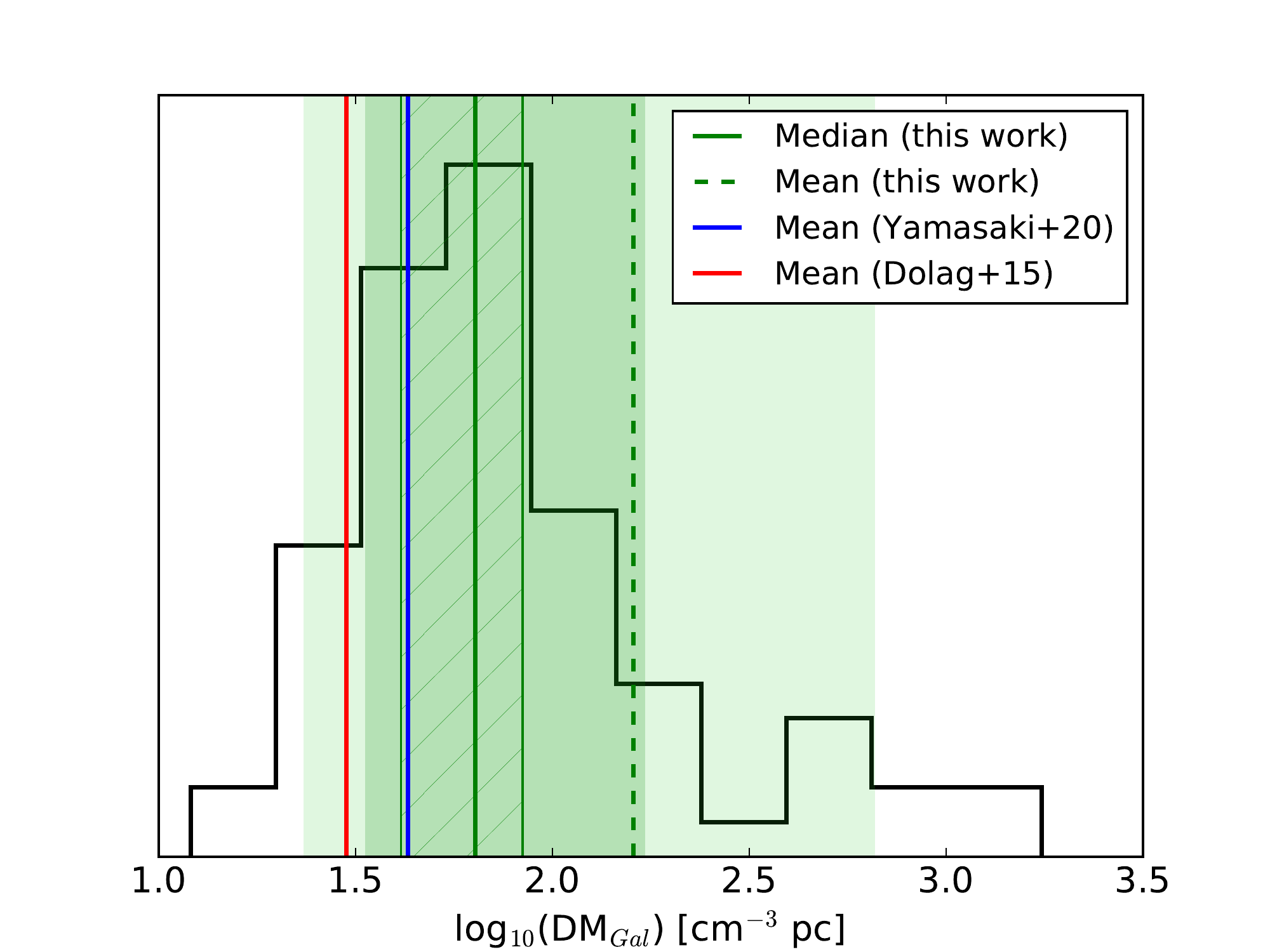}
    \caption{The histogram of Galactic dispersion measure in log$_{10}$ scale. The distribution is asymmetric, with a tail toward higher values. This is reflected by the stark difference between the median and the mean of the distribution. The hatched region corresponds to the uncertainty in the median. The dark (light) shaded region corresponds to 68\% (90\%) confidence interval. Overall, our estimate of Galactic DM is larger than the previous estimates. The mean of those estimates are shown for comparison.}
    \label{fig:stat}
\end{figure}

Next, we calculate the mean and the median of the DM$_{Gal}$ distribution. We do not include the lower limits and upper limits. The average DM$_{Gal}$ ranges from 12 to 1749 cm$^{-3}$ pc. This is larger than the ranges predicted by \cite{Prochaska2019} based on the absorption analysis (50--80 cm$^{-3}$ pc) and by \cite{Yamasaki2020} based on emission analysis (30--245 cm$^{-3}$ pc). There are only 8 out of 72 sightlines with the average DM$_{Gal} >$ 245 cm$^{-3}$ pc, and 9 sightlines with the average DM$_{Gal} <$ 30 cm$^{-3}$ pc. That means the DM$_{Gal}$ of most (76\%) of the sightlines are consistent with the estimate of \cite{Yamasaki2020}. The histogram of DM$_{Gal}$ is asymmetric toward the higher values (Figure \ref{fig:stat}), making the mean (161 cm$^{-3}$ pc) significantly higher than the median (64 cm$^{-3}$ pc). 

Please note that the DM$_{Gal}$ values are not as robust as the N(\oviin) values. DM$_{Gal}$ depends on both N(\oviin) and $f_{\hbox{\oviin}}$ (see equation 5). The uncertainty in the value of $f_{\hbox{\oviin}}$ depends on the robustness of temperature as well as the value of $f_{\hbox{\oviin}}$ at the temperature of the hot gas. The temperature depends on N(\oviiin) and N(\oviin), the error in both oxygen lines are propagated in the uncertainty of the temperature, making it less constrained than the individual lines. If $f_{\hbox{\oviin}}$ changes rapidly within the range of the temperature of the hot gas, the uncertainty in DM$_{Gal}$ will be driven by the uncertainty in $f_{\hbox{\oviin}}$, irrespective of how robust the \ovii (and \oviiin) measurement is. Secondly, the estimated DM$_{Gal}$ from multiple studies can be different due to the difference in method and/or instrument. This adds another uncertainty in DM$_{Gal}$ when the values are averaged.  

Keeping the above discussion in mind, we consider the distributions of DM$_{Gal} - \sigma_l$ and DM$_{Gal} + \sigma_u$. Here $\sigma_l$ and $\sigma_u$ are the statistical uncertainty of the average DM$_{Gal}$ in the lower and the upper end, respectively. The median of these two distributions provide an uncertainty in the median of the DM$_{Gal}$ distribution. We find that median $DM_{Gal} = 64^{+20}_{-23}$ cm$^{-3}$ pc (Figure \ref{fig:stat}). Additionally, we calculate the uncertainty in the mean by propagating the uncertainty of individual sightlines assuming Poissonian statistics, and obtain mean $DM_{Gal} = 161^{+243}_{-32}$ cm$^{-3}$ pc. The 68\% (90\%) confidence interval of the DM$_{Gal}$ distribution is 33--172 (23--660) cm$^{-3}$ pc. Our typical DM$_{Gal}$ is larger than the mean based on density models and cosmological hydrodynamic simulations \citep[43 and 30 cm$^{-3}$ pc, respectively;][]{Dolag2015,Yamasaki2020}. 

\subsection{Utility}
\noindent For an FRB localized in the sky, one needs to find the closest sightlines (i.e., the sightlines at smallest angular separation from the FRB) and obtain the mean of the Galactic DM along those sightlines. This would be the Galactic contribution to the total DM toward that FRB. The choice of the statistic (e.g., mean, median or interpolation) to combine multiple sightlines and the angular separation within which to consider the sightlines may vary with the scientific purpose. Using the DM$_{Gal}$ of a single sightline would be the simplest option, although that might not be the most accurate estimate. 

We attach a machine readable file with the paper. It has the Galactic longitude and latitude, average DM$_{Gal}$ and associated statistical and systematic uncertainty, and the maximum DM$_{Gal}$ along all sightlines considered here. The systematic uncertainty along a sightline reflects the scatter between the individual estimates along that sightline. Thus, for the sightlines measured once, there is no systematic uncertainty. Statistical uncertainty along a sightline is obtained by propagating the uncertainty in individual measurement along that sightline in quadrature. 

We build a code to extract the DM$_{Gal}$ toward an FRB. We attach a copy of the code with the paper. It takes the galactic coordinate of the FRB of consideration as system arguments in the units of degree, reads the galactic coordinates of the 72 sightlines from the file mentioned in the previous paragraph, calculates the angular distance between the FRB and all the sightlines and returns the best estimate of Galactic DM and associated error using the following 3 methods. \\
I) The DM$_{Gal}$ and its statistical uncertainty along the sightline at smallest angular separation from the FRB. \\
II) The mean and the median of the sightlines within a threshold of angular separation from the FRB. \\
III) The mean and the median of the sightlines separated from the closest sightline within a tolerance limit. \\
The threshold and the tolerance of angular separation in method II and III are taken as system arguments in the units of degree. We recommend the user to try different values of threshold and tolerance instead of fixed preconceived values, and wisely choose the statistic and the method depending on the scientific interest. This code can easily be combined with ISM dominated models \citep[e.g.,][]{Cordes2002}. It can also be extended to compare with the DM values from previous studies \citep{Prochaska2019,Yamasaki2020}. If a nearby sightline within the user's choice of threshold is not found, our code prompts a message asking to use the median of our DM$_{Gal}$ distribution as an alternative to the earlier estimates from density models. 

\textcolor{black}{It should be noted that while the previous density models are all-sky by construction, they are constrained by observations of similar sky coverage as ours. That means, the models and our empirical table have similar amount of observational information. The DM$_{Gal}$ values toward the directions without any observational information are interpolated in the models assuming some geometric symmetry. Therefore, if there is no sightline in our database within the angular separation threshold of the user, we would recommend the following paths:\\ 
1) Calculate the DM$_{Gal}$ from the density models toward the FRB, and also along the closest sightline to the FRB, in our database (the output of method\,I in the previous paragraph). Then, scale it with our estimate toward the closest sightline, i.e., $\hbox{DM (our estimate toward the FRB)} = \frac{\hbox{DM (model estimate toward the FRB)}}{\hbox{DM (model estimate toward the closest sightline})}\times$ DM (our estimate toward the closest sightline) \\
2) Modify the angular separation threshold and re-calculate DM$_{Gal}$ from our table (the output of method\,II in the previous paragraph). Calculate the DM$_{Gal}$ from the density models along the sightlines within the new threshold, and use the same statistic of median/mean as for our empirical estimate. Then, use the similar scaling as shown in the previous point, by replacing the closest sightline with the sightlines within the threshold. We have tested this for a few FRBs and found that at $|b|\geqslant20^\circ$, where the halo dominates over the ISM, setting a threshold of upto 20--25$\deg$ produces a meaningful result.}

Please note that the DM from cool and warm phases are not included in this code (see \S\ref{sec:analys} for details). If needed, one can add the median DM from these phases to the output of the code, although the resulting correction will be minimal. 


\textcolor{black}{As an example, we discuss the case of FRB\,20200125A at ($l,b$) = (359.8,58.4) \citep{Parent2020}. It has a DM of 179 cm$^{-3}$pc. For the ISM contribution to the DM being $\approx25$ cm$^{-3}$pc \citep{Cordes2002}, and by assuming no contribution from the Galactic halo,  the maximum redshift of the host is $z_{max}$ = 0.17. Assuming a fiducial Galactic halo contribution of 50 cm$^{-3}$pc, it reduces to $z_{max}$ = 0.12. The nearest sightline in our database, which is 7.3$\deg$ away from the direction toward the FRB has DM$_{Gal}$ = 645$_{-459}^{+3148}$ cm$^{-3}$pc. Because this is not constraining enough for this particular FRB, we consider the median of all sightlines within 20$\deg$ of the FRB, and find DM$_{Gal}$ = 71$\pm$7 cm$^{-3}$pc, including all phases. Similar exercise for the density models of \cite{Prochaska2017} and \cite{Yamasaki2020} predicts 39 and 41 cm$^{-3}$pc, respectively. These are smaller than our prediction by 4$\sigma$. The  predicted redshifts of the FRB based on the density models are larger than that based on our empirical estimate by 37\%. Once the redshift of the host of this FRB is determined, the density models and our empirical results can be tested to see which one is a better description of the Galactic halo.}

\section{Conclusions}\label{sec:summary}
\noindent Based on 21-cm, UV and X-ray absorption analyses at $z=0$ along 72 sightlines, we provide an empirical list of the Galactic dispersion measure contribution to the extragalactic fast radio bursts. It is independent of any density model and the spatial extent of the Galactic halo. We improve upon the measurements of N(\oviin) and f$_{OVII}$ compared to previous studies, thus providing a better estimate of DM in the hot phase. Our findings are: \\
1) $DM_{Gal}$ is dominated by the hot phase probed by X-ray absorption.\\
2) There is no definite trend of $DM_{Gal}$ with respect to the galactic coordinates. \\
3) The previous models on average are consistent with our measurements, but there is significant scatter around the average. There are also a few sightlines where our measurements are significantly different. \\
4) The median DM$_{Gal}$=64$^{+20}_{-23}$ cm$^{-3}$ pc. The 68\%(90\%) confidence interval is 33--172 (23--660) cm$^{-3}$ pc.   \\

Different methods of DM estimates have their own strengths and weaknesses. The disk-plus-halo models based on X-ray emission make a lot of assumptions, but the models can provide a rough estimate of the DM along any sky direction. Absorption measurements have significantly smaller sky coverage, but they have significantly fewer assumptions compared to emission based studies; this way they are more accurate. The work we present here is complementary to previous works, and provides more accurate DM values along the sightlines probed. We provide a table and a code to retrieve $DM_{Gal}$ for any FRB localized in the sky.

\section*{Acknowledgements}
SM and SD gratefully acknowledges the support from the NASA grant NNX16AF49G.  YK acknowledges support from grant DGAPA-PAPIIT 106518, and from program DGAPA-PASPA.
\section*{Data Availability}
The data behind figures are tabulated in \ref{tab:gupta},\ref{tab:nicastro},\ref{tab:gatuzz} and \ref{tab:all}.

\bibliographystyle{mnras}



\appendix
\section{Tables}
\begin{table}
\centering
\caption{DM$_{hot}$ based on the \ovii and \oviii measurements of $\hbox{\protect \cite{Gupta2012,Gupta2017}}$ and DM$_{cold}$ based on the 21-cm measurements of $\hbox{\protect \cite{Bekhti2016}}$. The last two columns denote the uncertainty in DM$_{hot}$. (Full table is available online)}
\resizebox{0.5\textwidth}{!}{%
\begin{tabular}{ccccccc}
\label{tab:gupta}\\
\toprule
Target      & l      & b      & DM\_cold & DM\_hot & e\_low   & e\_up  \\
            & [deg] & [deg]  & [cm$^{-3}$pc] & [cm$^{-3}$pc] & [cm$^{-3}$pc] & [cm$^{-3}$pc] \\
\midrule
\multicolumn{7}{c}{Both \ovii and \oviii available:}\\
3C382       & 61.30  & 17.44  & 4.01 & 98.01  & 52.02  & 52.07 \\
ARK564      & 92.13  & -25.33 & 3.23 & 43.86  & 34.40  & 20.12 \\
Mrk 509$^a$    & 35.97  & -29.86 & 2.56     & 156.61  & 48.98  & 48.25 \\
\multicolumn{7}{c}{Only \ovii available. $f_{OVII} = 1 $ :}\\
1ES1927+654 & 96.98  & 20.96  & 4.16 & 41.95 & 16.75 & 16.75 \\
3C273       & 289.95 & 64.36  & 1.09 & 22.53 & 6.07  & 6.07  \\
\bottomrule
\multicolumn{7}{c}{$^a$From hybrid ionization modeling by \cite{Gupta2017}, DM$_{hot} = 89.16^{+47.78}_{-29.94}$ cm$^{-3}$pc} 
\end{tabular}%
}
\end{table}

\begin{table}
\centering
\caption{DM$_{hot}$ based on the \ovii and \oviii measurements of $\hbox{\protect \cite{Nicastro2016a}}$ and DM$_{cold}$ based on the 21-cm measurements of $\hbox{\protect \cite{Bekhti2016}}$. The last two columns denote the uncertainty in DM$_{hot}$. (Full table is available online)}
\resizebox{0.5\textwidth}{!}{%
\begin{tabular}{ccccccc}
 \label{tab:nicastro}\\
\toprule
Target            & l      & b      & DM\_cold   & DM\_hot & e\_low    & e\_up     \\
& [deg] & [deg]  & [cm$^{-3}$pc] & [cm$^{-3}$pc]  & [cm$^{-3}$pc] & [cm$^{-3}$pc] \\
\midrule
\multicolumn{7}{c}{Both \ovii and \oviii available:}\\
SAXJ1808.4-3658   & 355.39 & -8.15  & 7.78  & 63.79   & 12.48   & 13.23   \\
XTEJ1650-500      & 336.72 & -3.43  & 39.56 & 179.79  & 33.71   & 481.70  \\
\multicolumn{7}{c}{Only \ovii available. $f_{OVII} = 1 $ :}\\
PSRB0833-45  & 263.55 & -2.79  & 1.95  & 31.76   & 16.35   & 39.89    \\
\multicolumn{7}{c}{\ovii and 3$\sigma$ upper limit of \oviii available:}\\
3C120          & 190.37 & -27.40 & 6.49  & $<$113.02    \\
\multicolumn{7}{c}{$f_{OVII} = 1 $, neglecting the upper limit of \oviiin:}\\
MAXIJ0556-332  & 238.94 & -25.18 & 2.59  & 46.32   & 22.78   & 35.75    \\
\bottomrule
\end{tabular}
}
\end{table}

\begin{table}
\centering
\caption{ DM$_{cold}$ and DM$_{hot}$ based on the ionization modeling  of $\hbox{\protect\cite{Gatuzz2018}}$ (Full table is available online) }
\resizebox{0.5\textwidth}{!}{%
\begin{tabular}{ccccccc}
\label{tab:gatuzz}\\
\toprule
Target            & l      & b      & DM\_cold & err  & DM\_hot & err     \\
& [deg] & [deg]  & [cm$^{-3}$pc] & [cm$^{-3}$pc] & [cm$^{-3}$pc] & [cm$^{-3}$pc] 
  \\ \midrule
4U 1254–69          & 303.48 & -6.42  & 20.75 & 0.06 & 39.96 & 6.32 \\
4U 1543–62          & 321.76 & -6.34  & 15.37 & 0.45 & 21.13 & 11.97 \\
4U 1636–53          & 332.91 & -4.82  & 24.97 & 0.65 & 32.45 & 13.19 \\
4U 1735–44          & 346.05 & -6.99  & 23.67 & 0.71 & 35.45 & 13.70 \\
4U 1820–30          & 2.79   & -7.91  & 6.49  & 0.39 & 24.28 & 5.31  \\
\bottomrule
\end{tabular}
}
\end{table}

\begin{table}
\centering
\caption{Average and maximum Galactic dispersion measure, combining multiple methods and instruments.(Full table is available online)}
\resizebox{0.5\textwidth}{!}{%
\begin{tabular}{cccccccc}
\label{tab:all}\\
\toprule
Target            & l      & b      & DM$_{avg}$ & e\_low &  e\_up  & DM$_{max}$ & e\_sys     \\
& [deg] & [deg]  & [cm$^{-3}$pc] & [cm$^{-3}$pc] & [cm$^{-3}$pc] & [cm$^{-3}$pc] & [cm$^{-3}$pc] 
\\
\midrule 
\multicolumn{8}{c}{one dataset}             \\
4U 1254–69          & 303.48     & -6.42      & 60.71       & 6.32          & 6.32     & 67.04    &  ---    \\
4U 2129+12           & 65.01      & -27.31     & 21.07       & 13.63         & 31.93    & 53.00    &  ---    \\
\multicolumn{8}{c}{combined}             \\
4U 1543–62          & 321.76     & -6.34      & 69.26       & 6.59          & 148.45   & 465.46   & 32.77   \\
4U 1636–53          & 332.91     & -4.82      & 92.63       & 10.08         & 14.78    & 161.55   & 35.22   \\
4U 1735–44          & 346.05     & -6.99      & 94.22       & 7.87          & 11.65    & 154.34   & 35.10   \\
\bottomrule
\end{tabular}
}
\end{table}


\bsp	
\label{lastpage}
\end{document}